\documentclass[aps,prl,twocolumn,superscriptaddress,showpacs]{revtex4}
\usepackage{epsfig}
\usepackage{amsmath}
\usepackage{times}

\begin{document}

\title{Heat conduction in simple networks: Controlling heat flow through inter-chain
coupling}

\author{Zonghua Liu}
\affiliation{Institute of theoretical physics and Department of
Physics, East China Normal University, Shanghai, 200062, China}
\author{Baowen Li}
\affiliation{Department of Physics and Centre for Computational
Science and Engineering, National University of Singapore, 117542
Singapore}

\date{22 March 2007}

\begin{abstract}
The heat conduction in simple networks consisting of different one
dimensional nonlinear chains is studied. We find that the coupling
between chains has different function in heat conduction compared
with that in electric current. This might find application in
controlling heat flow in complex networks.

\end{abstract}

\maketitle

Energy and information transports on networks, such as the
metabolism network, neuronal network, porous material network, and
oil production network, etc., have been studied for a long time
\cite{Stauffer:1992} and is recently getting more attention
because of the hectic activity in the complex networks and the
great progress in nanoscale fabrication technology where the
naotube/nanowire networks can be made for different purposes
\cite{Yakov:2004,Kumar:2005,Chub:2005,Lopez:2005,Wu:2005,Lizana:2005,Albert:2002}.
It is found that the electric transport changes linearly with the
number of added bonds \cite{Chub:2005,Kumar:2005}. The whole
resistance of network can be figured out by the Kirchhoff second
law for the complicated parallel and serial electric circuit
\cite{Lopez:2005}.

However, little is known about heat conduction in the complex
networks, although some progress has been achieved in the study of
heat conduction in single one dimensional chains (See
Ref.\cite{Review} and the references therein). The fundamental
question for heat conduction in one dimensional chains is that
what is the necessary and/or sufficient condition for the heat
conduction to obey the Fourier law. From computer simulations, it
is found that in 1D nonlinear lattices with on-site potential such
as the Frenkel-Kontorova (FK) model and the $\phi^4$ model, the
heat conduction obeys the Fourier's law, namely, the heat
conductivity is size independent\cite{FK}, which is  also called
normal heat conduction. Whereas in other nonlinear lattices
without on-site potential, thus momentum is conserved, such as the
Fermi-Pasta-Ulam (FPU) and alike models, the heat conduction
exhibits anomalous behavior\cite{FPU}, namely the heat
conductivity $\kappa$ diverges with the system size $N$ as $\kappa
\sim N^{\delta}$. A great effort has been devoted to understand
the physical origin and the value of the divergent exponent
$\delta$\cite{Exponent}. It is found that the anomalous heat
conduction is due to the anomalous diffusion and a quantitative
connection between them has been established\cite{Li}. Most
recently, we found that both the normal and anomalous heat
conduction can be described by an effective phonon theory under
the same framework\cite{LTL}.

More importantly, it is found that a single 1D chain consists of
two different lattices exhibit very interesting physical phenomena
such as thermal rectification\cite{Diode}  and negative
differential thermal resistance\cite{LWC2}. Experimental on
nanotube has verified the rectification \cite{Berkely}. Opening up
a new field of controlling heat flow from simple (or complex) nano
scale networks.

All studies on the 1D single chain can be regarded as the first
step to understand the heat conduction on realistic situations,
i.e., complex networks. In general, a complex network consists of
many 1D (or quasi-1D) chains with a diversity of couplings among
them. Therefore, the key to understand the heat conduction on
networks is to understand the influence of coupling to heat fluxes
in simple networks, i.e., coupled chains. According to the best of
our knowledge, this problem has not been investigated so far.

For the sake of simplicity, we would like to consider $m$ 1D
chains with several couplings between any two of them. To be more
specific, we take the FPU-$\beta$ chain\cite{FPU} as the basic
element and each chain are contacted with the Nose-Hoover
thermostat \cite{Nose:1984} at the two ends, keeping the first and
the last particle of the chain at temperature $T_h$ and $T_l$,
respectively. Without coupling, each chain has a Hamiltonian
 $H=\sum_i\frac{1}{2}
 p_i^2+V(x_i,x_{i+1})$, where
 $V(x_i,x_{i+1})=\frac{1}{2}\sum_i(x_{i+1}-x_i)^2+\frac{1}{4}\sum_i(x_{i+1}-x_i)^4$,
$x_i$ represents the displacement from the equilibrium position of
the i'th particle. The motion of the particles for
$i=2,3,\cdots,N-1$ satisfy the canonical equations $
\dot{x_i}=\frac{\partial H}{\partial p_i}$;
$\dot{p_i}=-\frac{\partial H}{\partial x_i}$. The dynamical
equations for the heat baths are $
\dot{\xi}_{h}=\frac{\dot{x}_1^2}{T_h}-1,
\dot{\xi}_{l}=\frac{\dot{x}_N^2}{T_l}-1$. The dynamical equations
for the first and last particles are $ \dot{p}_1=-\frac{\partial
H}{\partial x_1}-\xi_h p_1$, $\dot{p}_N=-\frac{\partial
H}{\partial x_N}-\xi_l p_N$.

The temperature is defined as $T(i)=\langle p_i^2\rangle$ and the
heat flux along the chain is $J=\langle p_i\frac{\partial
V}{\partial x_{i+1}}\rangle$. Suppose there is a coupling between
the node $i$ of one chain and the node $j$ of another chain, then
we have an additional new potential $V_{ij}'=\frac{1}{2}(x_{i}-
x_j)^2+\frac{1}{4}(x_{i}-x_j)^4$. The equations of node $i$ and
node $j$ become
\begin{equation}\label{eq:coupling}
\dot{p_i}=-\frac{\partial H}{\partial x_i}-\frac{\partial
V_{ij}'}{\partial x_i}; \dot{p_j}=-\frac{\partial H}{\partial
x_j}-\frac{\partial V_{ij}'}{\partial x_j}.
\end{equation}

To investigate the influence of coupling by numerical simulations,
we take $T_h=0.7$ and $T_l=0.5$ for all the chains and first
consider the case of $m=2$, i.e., two coupled chains in this
Letter. The two chains are coupled at different nodes $i,j$. We
find that both the temperature distribution and the total flux in
the steady state are changed with the coupling positions.

{\bf Case I: two chains of the same length  coupled together}

\begin{figure}
\epsfig{figure=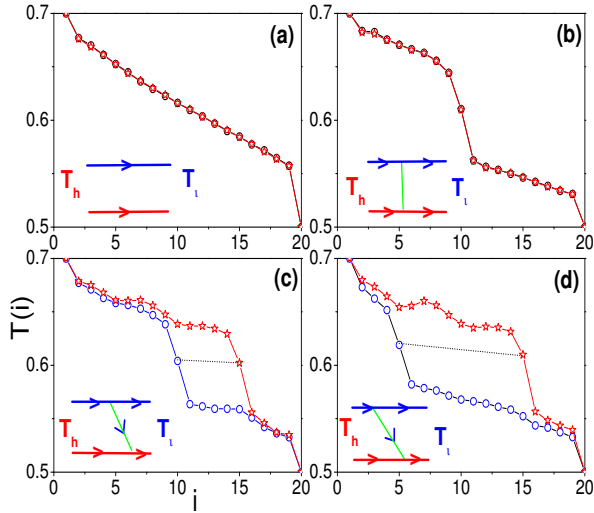,width=\linewidth,height=8cm}
\vspace{-1.2cm} \caption{Temperature distributions of two coupled
chains of length $N=20$ with different coupling positions. The
thin lines (green) denote the coupling. The insets are the
schematic configurations of coupled chains, and the arrows there
label the direction of heat flow. (a) No coupling, (b) Coupling at
$i=j=10$. (c) Coupling at $i=10,j=15$. (d) Coupling at $i=5,j=15$.
} \label{2chain_1}
\end{figure}

 The two chains are identical of length $N=20$. Like all models of
 heat conduction, there are always temperature jumps at the two
 boundaries\cite{Jump} as is clearly shown in Fig. \ref{2chain_1}.
 When two chains are coupled together
regardless of the coupling position, there is also temperature
jump at the junction, see Fig. \ref{2chain_1}(b)-(d).

\begin{figure}
 \epsfig{figure=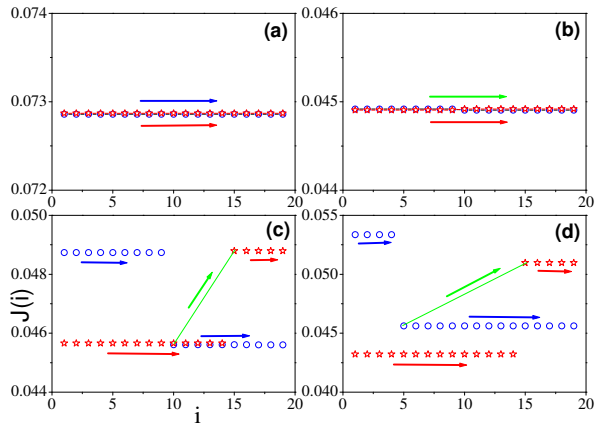,width=\linewidth}
 \vspace{-1cm} \caption{ The
corresponding fluxes of Fig. \ref{2chain_1}.}
 \label{2chain_2}
\end{figure}

In Fig. \ref{2chain_2} where we shows the corresponding fluxes of
Fig. \ref{2chain_1} where the arrows denote the directions of
fluxes. Fig. \ref{2chain_2}(a) is easy to understand from their
identity, where two uncoupled chains have the same flux.

Fig. \ref{2chain_2}(b) shows a very interesting result -  the
reduction of the heat current. This is completely different from
 electric circuit. It is well known that a circuit of
two chains with four equal resistance R connected by a conduction
line at the middle is a symmetric circuit. Since there is no
potential difference between the two connecting points, there is
no current through the middle connection line, thus the current in
the circuit does not change! It remains the same if the two chains
are disconnected.

What makes the "thermal circuit" different from the electric
circuit? To this end, we need to go to the definition of
temperature. The temperature is a measure of the kinetics of the
particle. It is an ensemble (time) average of the kinetic energy.
Without coupling, the middle particle at each chain is connected
only by its two nearest neighbors. After coupling, the middle
particle is connected with three particles which changes its
equation of motion. Even though the two particles in the middle
have the same temperature (same average kinetic energy and same
velocity distribution), it does not mean that the two particles
always oscillate in the same way. This is the fundamental
difference between the electric circuit and thermal circuit.

In fact, the coupling of the second chain to the first chain is
equivalent to the introduction of an interface resistance at the
junction. This resistance is also called the Kapitza resistance
which is defined as: $R_{int}=\Delta T/J$, where the $\Delta T$ is
the temperature jump between the left and right particles of the
interface (coupled particle in the middle). Therefore the heat
current through each chain is:
\begin{equation}
J=\frac{T_h-T_l}{2R+R_{int}}
\end{equation}
which is obviously less than $J_0=\frac{T_h-T_l}{2R}$ for the
uncoupled chain.

In the case of without any coupling, the temperature of the $i'th$
particle inside the FPU chain is:
\begin{equation}
T_i\approx T_h-|\Delta T_h|-\frac{i-1}{N-2}\left(T_h-|\Delta
T_h|-T_l-|\Delta T_l|\right), \label{eq:Tofi}
\end{equation}
where $|\Delta T_h|$ and $|\Delta T_l|$ is the temperature jump at
the both ends between the heat bath and the first/last particle of
the chain, respectively. The heat current flows at the junction
can be understood from this formula. For instance, the particle at
$i=10$, has higher temperature than the particle of $i=15$. Heat
flows always from high temperature to low temperature, therefore,
if one connects $i=10$ in upper chain to particle $i=15$ in lower
chain, there will be heat current flows from $i=10$ (higher
temperature) in upper chain to particle $i=15$ (low temperature)
in the lower chain. This will drag the temperature of particle
$i=10$ down a little bit, thus we see the increase of the heat
current in the part of $i\in [2,10]$ in upper chain in Fig. 2(c)
compared with the case in Fig. 2(b). In contrast, as the heat
current flows to particle $i=15$ at lower chain, the temperature
at $i=15$ is increased, thus the increase of the temperature
difference between $i=15$ and $i=20$, which leads to the increase
of heat current in segment of $i\in[15,20] $ in lower chain. This
is what we observe in Fig. \ref{2chain_1} (c). The same mechanism
applies also to Fig \ref{2chain_1} (d).

{\bf Case II: Two chains of different length  coupled together}

\begin{figure}
\epsfig{figure=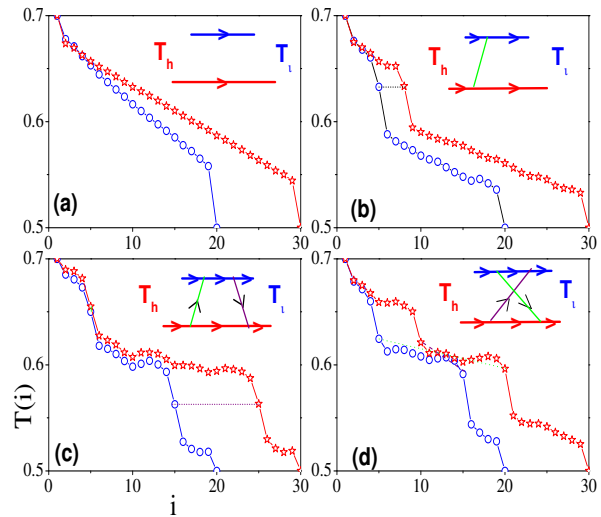,width=\linewidth,height=8cm}\vspace{-1cm}
\caption{Temperature distributions of two coupled chains with
different lengthes $N_1=20$ and $N_2=30$ with different coupling
positions. The thin (green and purple) lines denote the coupling.
The insets are the schematic configurations of coupling chains,
and the arrows indicate the direction of heat flow. (a) no
coupling, (b) one coupling added at $i=5, j=8$; (c) two couplings
added at $i_1=5, j_1=5$ and $i_2=15, j_2=25$; (d) two crossing
couplings added at $i_1=5, j_1=20$ and $i_2=15, j_2=10$. }
\label{2chain_3}
\end{figure}

We would like to extend above ideas to more general case, namely,
 coupling of two chains of different length and multiple coupling.
 We find that the reduction of heat flux by coupling is quite
general. Fig. \ref{2chain_3} shows the temperature distribution of
two coupled chains with different lengthes $N_1=20$ and $N_2=30$,
respectively. (See Figure caption for more information.) It is
easy to see that Fig. \ref{2chain_3} has some similarity with Fig.
\ref{2chain_1}, i.e., there are temperature jumps at the coupled
particles and the coupled particles have the approximate same
temperature.

Another interesting thing is that the crossing couplings make the
middle part of the coupled chains appear a temperature plateau
which might be useful in heat control.

\begin{figure}
\epsfig{figure=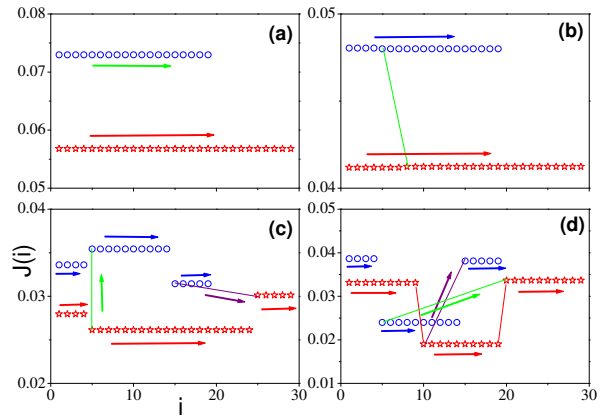,width=\linewidth} \vspace{-1cm}
\caption{ The corresponding fluxes of Fig. \ref{2chain_3}. }
\label{2chain_4}
\end{figure}

Fig. \ref{2chain_4} shows the corresponding fluxes of Fig.
\ref{2chain_3}. The longer chain, $N_2=30$ has smaller heat
current. Although the coupling of the two chains of different
length is not at the symmetrical point, there is still no current
through the coupling.

 In fact, the like in the previous case shown in Fig. (1) and (2), the
 heat current flow in the (multi) coupled chain of different length
 can be also understood from Eq. (\ref{eq:Tofi}).
 According to this formula, we can roughly
estimate that the $T_5$ at short chain is roughly the same as
$T_8$ in the longer chain. Thus there is no heat flow between
them. The only influence is the introduction of an interface
resistance which drags down the heat current through each chain as
is seen in Fig. \ref{2chain_3}(b).

It is not difficult to estimate that $T_5$ at chain $N=30$ is
larger than $T_5$ at chain $N=30$. This is why we see the current
flow from particle 5 at lower (longer) chain to particle 5 at
upper (shorter) chain in Fig. \ref{2chain_4}. However, the
temperature of particle 15 at upper chain (shorter) is almost the
same as the temperature of particle 25 at lower chain (longer) if
the two chains are uncoupled. However, due to first coupling, the
temperature of particle 15 at upper chain is slightly increased,
this is why we see the current flows from particle 15 in upper
chain to the particle 25 in low chain as shown in Fig
\ref{2chain_4}(c).

More complicated and more interesting case is shown in Fig.
\ref{2chain_4}(d), where we have two crossing couplings: $i_1=5$
is connected to $i_2=15$, and $j_1=20$ is connected to $j_2=10$.
Use Eq.(\ref{eq:Tofi}), we can again estimate that $T_{i_1=5}>
T_{i_2=15}$, thus we see the current flow from upper chain to the
lower chain. Similarly, there is heat current flows from lower
chain (particle $j_2=10$) to upper chain ($j_1=20$).

We have also checked the heat conduction in multiple coupled
chains with a diversity of couplings, such as in the three coupled
chains with different lengthes, and observed the similar results
as in the case of two coupled chains. We conclude that, in
general, the coupling will introduce an interface resistance at
the junction thus affect the heat flow through the whole system.

{\bf Case III: Single chain with loop}

\begin{figure}
\epsfig{figure=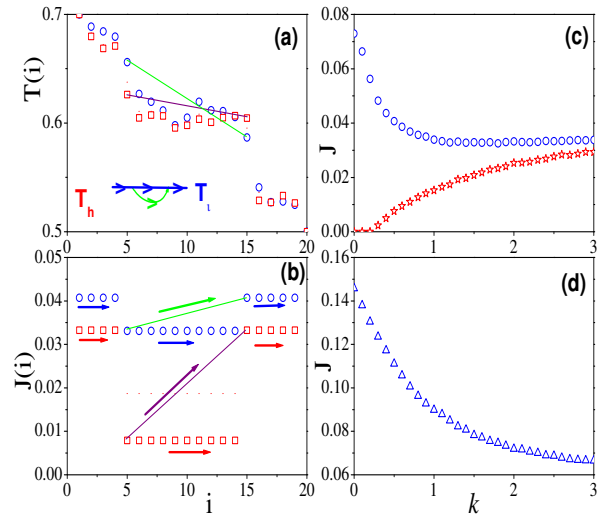,width=\linewidth,height=8cm}\vspace{-1cm}
\caption{(a) and (b) represent the distributions of temperature
and flux in a single chain with $N=20, i=5$ and $j=15$,
respectively, where ``circles" and ``squares" denote the cases of
$k=0.5$ and $2.0$, respectively. The arrows in the inset of (a)
indicate the direction of heat flow. (c) shows how the flux in (b)
changes with the coupling strength $k$ where the ``solid square"
are the total flux and the ``open circle" the flux through the
shortcut. (d) The heat current in Fig. \ref{2chain_2}(b) changes
with the coupling strength $k$.} \label{1chain}
\end{figure}

Another interesting question is how the self-coupling or a
shortcut in a single chain affect the heat current? This case
happens very frequently in the polymer chain and biological
systems???. For example, if there is a shortcut between the node
$i$ and the node $j$ of a chain (see the inset of Fig.
\ref{1chain}(a)), does this shortcut reduce the flux of the chain?
In the case of traffic flow (reference ????), the shortcut
increases the capacity of traffic because the vehicles have more
free space to go. However, in the thermal circuit, the flux should
be reduced because of the interface resistance. The line with
``stars" in Fig. \ref{1chain}(b) shows the result with $N=20,
i=5$, and $j=15$. Comparing it with Fig. \ref{2chain_2}(a) of no
coupling, it is easy to see that the flux is reduced almost
$50\%$.

We also study the dependence of the reduction of the heat flux on
the coupling strength $k$. From the definition of junction
resistance we know that the larger the degree of destroying the
correlation between the coupled particle and its neighbors is, the
larger $R_{int}$, i.e., $R_{int}$ should monotonously increase
with $k$. When the correlation is completely destroyed, $R_{int}$
cannot be increased more by further increasing $k$. Therefore,
there is a saturation effect for $R_{int}$ and the effect of
junction resistance when $k$ is large enough. Let's confirm this
prediction by numerical simulations. In this situation, the
coupling potential becomes
\begin{equation}\label{eq:potential}
 V_{ij}'=\frac{k}{2}(x_{i}-
x_j)^2+\frac{k}{4}(x_{i}-x_j)^4.
\end{equation}
Substituting Eq. (\ref{eq:potential}) into Eq. (\ref{eq:coupling})
we get the dynamical equations for the particles with coupling
strength $k$. Our numerical simulations show that for a single
chain with the self-coupling, the larger the coupling is, the more
reduction of flux. Fig. \ref{1chain}(b) shows three typical cases
where the lines with ``circles", ``stars" and ``squares" denote
the cases of $k=0.5$ and $2.0$, respectively. From the middle
parts of this figure it is ease to see that the larger coupling
makes less flux go through the original path. We notice that the
coupling also changes the temperature distribution. The strong the
coupling is, the two particles connected by the coupling have more
close temperatures, as shown in Fig. \ref{1chain}(a). For
observing the influence of coupling strength in more detail, Fig.
\ref{1chain}(c) shows how the fluxes change with the coupling
strength $k$ where the line with ``circles" denotes the total flux
and the line with ``stars" the flux going through the shortcut.
Obviously, the total flux becomes stabilized when $k>1$ and the
flux through the shortcut is monotonously increase with $k$,
confirming the saturation effect. The saturation effect has been
also observed in the coupling of two coupled chains, see Fig.
\ref{1chain}(d) for how the total flux of the two chains in Fig.
\ref{2chain_2}(b) changes with the coupling strength $k$.

In conclusions, we have studied the influence coupling in simple
networks on the heat conduction.  It is found that different from
the electric circuit, the coupling affect very much the heat
current flow in the thermal circuit. Any introduction of coupling
is equivalent to an introduction of an interface resistance, thus
influence largely the heat current in the circuit. The study may
shed lights for studying heat conduction in complex networks.

ZL is supported in part by the NNSF of China under Grant No.
10475027 and No. 10635040, by the PPS under Grant No. 05PJ14036,
by SPS under Grant No. 05SG27, and by NCET-05-0424. BL is
supported partially by a FRG grant and a ARF grant.

\end{document}